\documentclass[lettersize,journal]{IEEEtran}
\usepackage{amsmath,amsfonts}
\usepackage{algorithmic}
\usepackage{algorithm}
\usepackage{array}
\usepackage[caption=false,font=normalsize,labelfont=sf,textfont=sf]{subfig}
\usepackage{textcomp}
\usepackage{stfloats}
\usepackage[hidelinks]{hyperref}
\usepackage{url}
\usepackage{verbatim}
\usepackage{booktabs}
\usepackage{graphicx}
\usepackage{cite}
\usepackage{orcidlink}

\begin{document}

\title{A Shallow Recurrent Decoder for Dynamic State Estimation with a Limited Number of PMUs in Power Systems}

\author{Andrea Pomarico\orcidlink{0009-0006-6677-6399}, Alberto Berizzi\orcidlink{0000-0002-2856-783X} and J. Nathan Kutz\orcidlink{0000-0002-6004-2275}.
}
\markboth{Journal of XXX}%
{Pomarico\MakeLowercase{\textit{et al.}}: SHRED in power systems}


\maketitle

\begin{abstract}
Dynamic State Estimation (DSE) will play a fundamental role in future power system operation by providing real-time estimates of the system state and enabling enhanced situational awareness. Existing DSE approaches are primarily based on Kalman filter variants or Machine Learning (ML) techniques. However, Kalman-based methods often suffer from high computational complexity, sensitivity to model inaccuracies, and performance degradation under strongly nonlinear operating conditions. Moreover, their effectiveness critically depends on the number and placement of measurements, since suboptimal PMU locations can reduce observability and even render state estimation infeasible. Machine learning approaches alleviate some of these limitations but typically require large amounts of training data and may struggle to generalize.\\
To address these challenges, this paper proposes a SHallow REcurrent Decoder (SHRED) architecture for full-state reconstruction of power systems from sparse measurements. Unlike conventional model-based estimators, the proposed approach does not rely on an accurate physical model and is largely insensitive to PMU placement, making it particularly attractive for practical deployment in existing Wide Area Measurement Systems (WAMS). The method is validated on the IEEE 39-bus system under strongly nonlinear conditions, including short-circuit disturbances. The results demonstrate that SHRED can accurately reconstruct the complete system state using only a limited number of PMU measurements, consistently outperforming a state-of-the-art shallow decoder benchmark in sparse-measurement scenarios. Furthermore, the proposed framework exhibits strong robustness to measurement noise and maintains high reconstruction accuracy even under severe disturbances, highlighting its potential as a scalable and reliable alternative to conventional DSE techniques.
\end{abstract}


\section{Introduction}
\IEEEPARstart{I}{n} recent years, modern power systems have undergone profound transformations driven by the large-scale integration of Renewable Energy Sources (RES) \cite{HASSAN2024100545}. These resources are predominantly interfaced through power electronic converters, whose dynamic behavior is significantly faster and fundamentally different from that of conventional synchronous generators historically dominating power system operation \cite{kundur2004definition,hatziargyriou2020definition}. 
Given the vast geographical extent and increasing complexity of interconnected power grids, the likelihood of faults, disturbances, and cascading failures has markedly increased. Under such conditions, unexpected events may propagate rapidly, potentially triggering large-scale blackouts, as occurred to the Iberian power system in April 2025 \cite{entsoe2025,11218068}. Beyond their harmful impact on grid stability and reliability, these events can also produce severe social and economic consequences for the affected populations \cite{gonzalez2025social}. These risks have prompted Transmission System Operators (TSOs) to widely deploy Phasor Measurement Units (PMUs) within Wide-Area Measurement Systems (WAMS), enabling the extraction of meaningful real-time insights from the large and continuous data streams characteristic of transmission networks \cite{monti2016phasor}.  
PMUs are critical components for the advanced monitoring of modern control strategies which are enabled by accurate Dynamic State Estimation (DSE).
%
In particular, DSE enables the real-time reconstruction of system operating states at high temporal resolution, thereby providing essential information for enhancing situational awareness and supporting fast, reliable protection and control actions in modern power systems \cite{zhao2019power,zhao2020roles}.  Improvements of DSE using limited PMUs, which is the focus of the current work using a SHallow REcurrent Decoder (SHRED) architecture, can help enhance the overall performance of modern power systems and enable robust control paradigms.\\
Power systems are difficult to model and emulate due to the vast spatial and temporal scales present in modern systems.  Further, they are large scale networks whose nonlinearly coupled components can induce instabilities, bifurcations and behaviors which are difficult to predict~\cite{porter2016dynamical}.  Some of the induced nonlinear coupling effects can be advantageous~\cite{proctor2005passive}, but most are difficult to characterize and  control~\cite{cornelius2013realistic,morrison2020nonlinear}.  Machine learning (ML) and Artificial Intelligence (AI) methods are now emerging as a paradigm~\cite{alimi2020review} for the monitoring, state-estimation and control of power systems, with a wide range of algorithms that can be deployed in practice~\cite{brunton2022data}.  ML and AI methods learn to exploit patterns of activity in the data, using supervised or unsupervised algorithms, in order to help with the various aspects of making robust power grids. Spatio-temporal methods are especially relevant in modern power grids, since their spatial domains are vast and the time-scales that need to be resolved critical for operation.  Moreover, power systems are typically only sparsely instrumented, with measurements available for only a limited subset of the state variables. This motivates the development of an appropriate deep learning architecture capable of reconstructing the full dynamic state from sparse and discrete measurements. As such, the SHRED architecture is a deep learning architecture based in sensing~\cite{williams2024sensing}, but which can be exploited to not only produce a DSE, but also construct stable and robust forecasts in a computationally efficient latent space \cite{gao2025sparse}.

\subsection{Bibliography Review and Goal}\label{sec:biblio}
State estimation in power systems can be broadly categorized into two main approaches: Static State Estimation (SSE) and DSE. SSE is a core tool for TSOs worldwide, enabling the computation of voltage magnitudes and angles across the network using non-synchronized measurements, such as those provided by SCADA systems \cite{d2016power,primadianto2016review}. SSE is based on an algebraic model and has been integrated into TSOs' energy management systems for decades, with well-established roles, implementation practices, and operational requirements.

In contrast, DSE relies on Differential–Algebraic Equations (DAEs)\footnote{Henceforth, bold uppercase symbols denote matrices, while bold lowercase symbols denote vectors.} given by
\begin{equation}\label{eq:DAE}
\begin{cases}
    \mathbf{\dot{x}}(t) = \mathbf{f}\big(\mathbf{x}(t), \mathbf{y}(t), \mathbf{u}(t), \mathbf{p}(t)\big), \\
    0 = \mathbf{g}\big(\mathbf{x}(t), \mathbf{y}(t), \mathbf{u}(t), \mathbf{p}(t)\big),
\end{cases}
\end{equation}
where $\mathbf{x}(t)$ denotes the state vector, $\mathbf{y}(t)$ the algebraic variable vector, $\mathbf{u}(t)$ the input vector, and $\mathbf{p}(t)$ the parameter vector. Functions $\mathbf{f}(\cdot)$ and $\mathbf{g}(\cdot)$ are generally nonlinear vector-valued functions describing the dynamic and algebraic behavior of the system, respectively.

In practical power system applications, measurements are acquired in a discrete-time manner. Therefore, the continuous-time DAE model in~\eqref{eq:DAE} must be discretized to be compatible with online measurements. Accordingly, the measurement equation at discrete time instant $k$ can be expressed as
\begin{equation}
    \mathbf{s}_k = \mathbf{h}\big(\mathbf{x}_k, \mathbf{y}_k, \mathbf{u}_k, \mathbf{p}_k\big),
\end{equation}
where $\mathbf{s}_k$ represents the measurement vector and $\mathbf{h}(\cdot)$ is a nonlinear measurement function. The goal of DSE is to track the dynamic state variables $\mathbf{x}(t)$ associated with synchronous generators, RES, and loads using measurements $\mathbf{s}_k$. By providing real-time information on these states, DSE enhances the capabilities of TSOs, supporting applications such as Dynamic Security Assessment (DSA), monitoring generator rotor angles and speeds during disturbances, and assessing potential voltage instability through the states of Automatic Voltage Regulators (AVRs). From a control perspective, DSE enhances situational awareness by improving the visibility of system operating conditions and enables the validation and calibration of control models \cite{8894035}. From a protection standpoint, several studies have demonstrated that DSE-based out-of-step protection schemes offer faster and more reliable relay actions compared to traditional protection approaches \cite{farantatos2015predictive, 7585093}.

In literature, several DSE algorithms have been proposed. Early work, such as \cite{huang2007feasibility}, investigated Kalman filter–based approaches \cite{shivakumar2008review,welch1995introduction,li2015kalman}. Due to the strong nonlinearities that characterize power system dynamics, multiple variants of the Kalman filter have been developed to improve robustness. In particular, the Extended Kalman Filter (EKF) has been applied to estimate the state variables of synchronous generators \cite{akhlaghi2016multi,zhao2016robust}. However, since the EKF relies on a first-order Taylor series approximation, it may suffer from large estimation errors or even divergence when the system exhibits strong nonlinear behavior, such as during severe disturbances \cite{wang2011alternative}. To overcome the limitations associated with linearization, more advanced nonlinear filtering techniques have been proposed for power system DSE, including the Unscented Kalman Filter (UKF) \cite{valverde2011unscented,qi2016dynamic,zhao2017robust,hou2022dynamic}, which was introduced to avoid linear approximation and improve accuracy on highly nonlinear systems. However, UKF-based methods require computational time and DSE algorithms needs to be faster than the sampling rate of measurements to keep in track with the real-time system dynamics. Furthermore, despite their improved performance, Kalman-type filters are known to operate reliably only under specific assumptions \cite{zhao2017dynamic}, which are often not fulfilled in practical power system applications, as demonstrated in \cite{wang2017assessing}.\\

All the aforementioned approaches are rooted in classical state estimation theory. However, in recent years, the rapid advancement of ML techniques has motivated the adoption of data-driven approaches across a wide range of engineering problems \cite{machlev2022explainable,li2024artificial,9743327}. In the context of power systems, ML has also been explored for state estimation applications. For instance, SSE has been addressed using Physics-Informed Graph Neural Networks in \cite{ngo2024physics,10408467}, while Physics-Informed Dynamic Mode Decomposition has been proposed for DSE in \cite{10667170}. Moreover, a hybrid-learning DSE (HL-DSE) framework for real-time estimation of just synchronous machine rotor angle and speed has been introduced in \cite{8998331}, neglecting all the other state variables. Although ML–based algorithms have demonstrated promising accuracy for DSE and often outperform classical model-based approaches, they still face significant challenges. These include high computational burdens during training, the need for large amounts of data, and extensive hyperparameter tuning, all of which can hinder their practical deployment in real-world power system applications.

\subsection{Paper Contribution}
The gaps identified in Section~\ref{sec:biblio} motivate the analysis carried out in this paper. The main contributions of this work are summarized as follows:
\begin{itemize}
    \item SHRED architecture is introduced for the first time in the power system domain.
    \item It is demonstrated that SHRED can accurately reconstruct the state variables of all synchronous generators, as well as the voltage magnitude, voltage angle and frequency at each bus of the power system, using only a limited number of PMU measurements.
    \item It is further shown that SHRED is largely agnostic to PMU placement, a property particularly desirable for practical DSE applications.
    \item It is demonstrated that SHRED is robust to measurement noise, which is known to characterize real-world data. Furthermore, ensembling multiple SHRED architectures provides more accurate state reconstruction than using any single model alone.
    \item An extensive performance assessment of SHRED for DSE is carried out on the IEEE 39-bus system, considering strongly non-linear events such as short-circuit faults.
\end{itemize}

The remainder of this paper is organized as follows. Section~\ref{sec:SHRED} introduces the SHRED architecture and reviews the related literature. Numerical simulation results are presented in Section~\ref{sec:results}. Limitations and future works are presented in Section \ref{sec:limitations}. Lastly, concluding remarks are drawn in Section~\ref{sec:conclusion}.

\section{SHallow REcurrent Decoder}\label{sec:SHRED}
\begin{figure*}
    \centering
    \includegraphics[width=.97\linewidth]{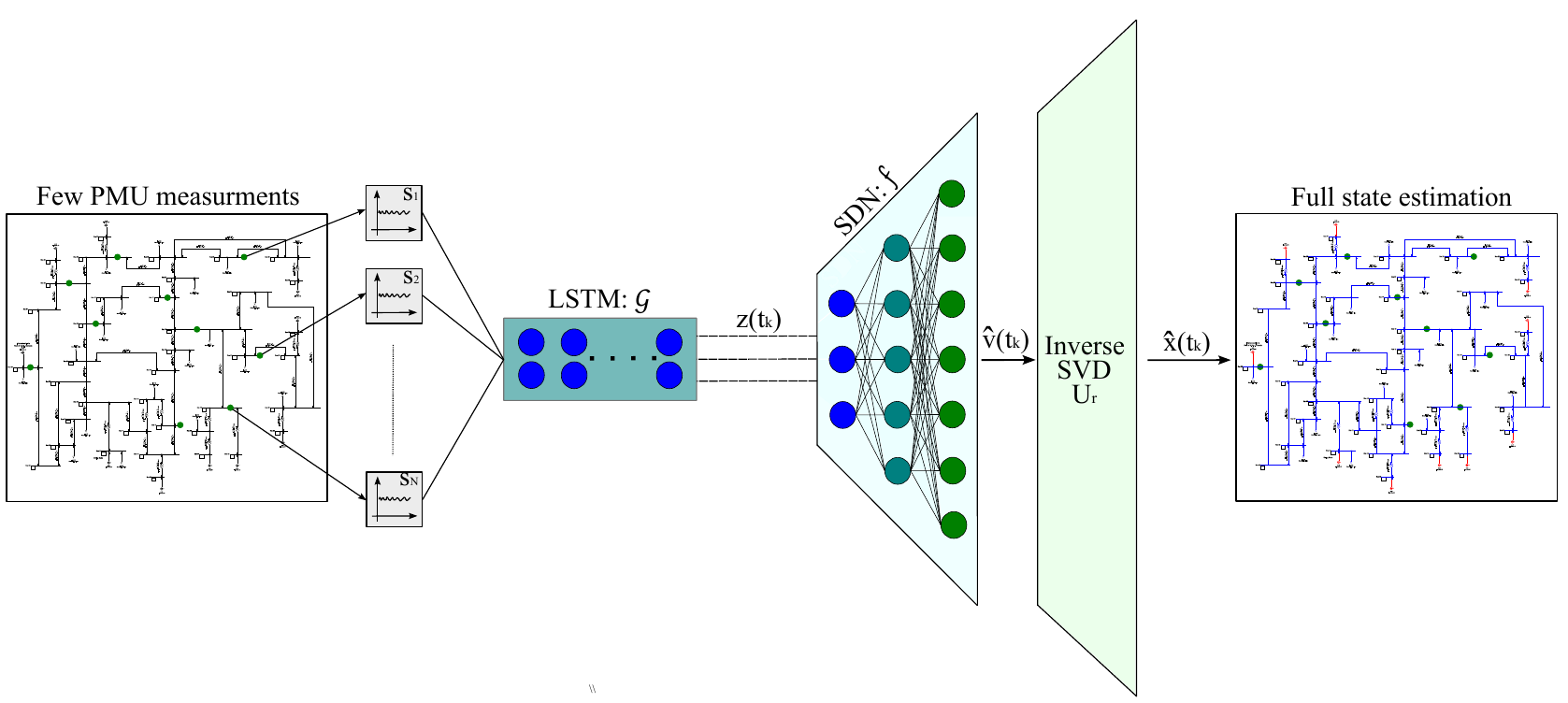}
    \caption{SHRED architecture applied to power systems. A limited number of PMUs are used to measure voltage and frequency signals. These sensor time series are processed to construct a latent temporal sequence model, which is then mapped to a compressed representation of the full system state. Finally, the compressed representation is projected back into the original state space through singular value decomposition.}
    \label{fig:SHRED}
\end{figure*}

The SHRED architecture, first introduced in \cite{williams2024sensing}, is a promising deep learning framework that has demonstrated outstanding reconstruction and forecasting capabilities, particularly in low-data regimes. Owing to its effectiveness, SHRED has achieved state-of-the-art performance across a wide range of scientific domains. For example, in \cite{faraji2025shallow}, the authors showed that SHRED can accurately reconstruct the full state space of plasma dynamics using only three measurement points. Similarly, SHRED has been extensively validated in the context of nuclear reactors \cite{riva2025robust, riva2025towards}, where it successfully reconstructs reactor dynamics from a limited number of out-of-core measurements.\\
In the biological domain, SHRED was applied in \cite{rude2025shallow} to reconstruct neural activity and behavior in mouse, zebrafish, and human. In \cite{tomasetto2025reduced}, it was employed for model order reduction, while in \cite{gao2025sparse}, SHRED was integrated with Sparse Identification of Nonlinear Dynamical Systems (SINDy) \cite{brunton2016discovering} and Koopman theory \cite{brunton2021modern} to discover spatiotemporal governing equations directly from data.\\
Finally, in \cite{bao2025data}, the Data Assimilation SHRED (DA-SHRED) framework was proposed to bridge the gap between reduced-order simulations and real-world physical systems characterized by high-dimensional spatiotemporal fields. By leveraging the latent space learned from reduced models and updating it with sparse real sensor measurements, DA-SHRED enables accurate reconstruction of the full system state even when direct observations are unavailable.\\
However, to date, SHRED has not been applied to power system dynamics, which are well known to be highly nonlinear and complex. In this work, we aim to demonstrate how SHRED can reconstruct the full state vector of a power system using only a few PMU measurements as input.
As a first step, SHRED is trained to map a limited number of measurements $\{s_1, \dots, s_N\}$, which, in the considered power system application, correspond to PMU signals, into a compressed space, obtained via a low-rank approximation using Singular Value Decomposition (SVD). This initial compression stage approximates the full-order snapshot space with a low-rank finite-dimensional subspace spanned by the leading left singular vectors of the SVD. As a result, the training cost is significantly reduced, enabling efficient training of the SHRED model within a few minutes even on a standard personal laptop. To illustrate this, Fig.~\ref{fig:SHRED} presents the SHRED architecture adapted for power system applications.
The basic SHRED architecture is composed of a Long Short-Term Memory (LSTM)\cite{graves2012long} and a Shallow Decoder Network (SDN)\cite{erichson2020shallow}. The LSTM constructs a latent representation of the underlying dynamics, a mechanism closely related to Takens’ embedding theorem \cite{takens2006detecting}. Then, the latent features are processed by SDN, which maps them to the compressed space of the full system. Finally, the reconstructed compressed states are projected back to the original high-dimensional state space via the inverse SVD transformation.

\subsection{Mathematical Background}
Let $\mathcal{G}_1$ denote the LSTM network that encodes the sequence of measurements $\mathbf{s}_i$ over the time window $[t_{k-l},t_k]$ into a latent representation $\mathbf{z}_k$ as
\begin{equation}
    \mathbf{z}_k(t_k) = \mathcal{G}_1\big(\mathbf{s}_i(t_k), \dots, \mathbf{s}_i(t_{k-l})\big),
\end{equation}
where $l$ represents the number of time lags considered. The full high-dimensional state vector $\mathbf{x}(t_k)$ is then reconstructed by the SDN decoder, represented by the function $\mathcal{F}_1$, which maps the latent space back to the full state space:
\begin{equation}
\tiny
\mathbf{\hat{x}}(t_k)=\mathcal{F}_1(\mathbf{z}(t_k))=\mathcal{F}_1\left(\mathcal{G}_1(\mathbf{s}_i(t_k), \dots, \mathbf{s}_i(t_{k-l})\right)
\normalsize
\end{equation}
here $\mathbf{\hat{x}}(t_k)$ is the reconstruction performed by SHRED of $\mathbf{x}(t_k)$. The recurrent encoder $\mathcal{G}_1$ and decoder $\mathcal{F}_1$ are trained jointly by minimizing the reconstruction error over the training dataset $\Xi_{\text{train}}$:
\begin{equation}
\begin{split}
\mathcal{J} &= \sum_{t_k \in \Xi_{\text{train}}} \left\| \mathbf{x}(t_k) - \mathbf{\hat{x}}(t_k) \right\|_2^2 \\
&= \sum_{t_k \in \Xi_{\text{train}}} \left\| \mathbf{x}(t_k) - 
\mathcal{F}_1\Big(\mathcal{G}_1(\mathbf{s}_i(t_k), \dots, \mathbf{s}_i(t_{k-l}))\Big) \right\|_2^2.
\end{split}
\end{equation}

To reduce redundancy and accelerate training, the high-dimensional state is projected onto a lower-dimensional subspace via the SVD. Let $\mathbf{X} \in \mathbb{R}^{n \times m}$ be the snapshot matrix collecting the $m$ full-state training vectors, with SVD $\mathbf{X} = \mathbf{U}\,\mathbf{\Sigma}\,\mathbf{V}^{\top}$, where the 
singular values are ordered as $\sigma_1 \geq \sigma_2 \geq \dots \geq 0$. Retaining the first $r \ll n$ left singular vectors $\mathbf{U}_r \in \mathbb{R}^{n \times r}$, during training the state is projected as
\begin{equation}
    \mathbf{v}(t_k) = \mathbf{U}_r^{\top}\,\mathbf{x}(t_k), 
    \label{eq:projection}
\end{equation}
The truncation rank $r$ is chosen to retain a fraction $\eta$ of the cumulative spectral energy,
\begin{equation}
    r = \min\Big\{ k : \textstyle
    \sum_{i=1}^{k} \sigma_i^{2} \big/ 
    \sum_{i} \sigma_i^{2} \geq \eta \Big\},
    \label{eq:rank}
\end{equation}
for $\eta = 99\%$. The training loss then becomes
\begin{equation}
\begin{split}
\mathcal{J} &= \sum_{t_k \in \Xi_{\text{train}}} \left\| \mathbf{v}(t_k) - \mathbf{\hat{v}}(t_k) \right\|_2^2 \\
&= \sum_{t_k \in \Xi_{\text{train}}} \left\| \mathbf{U}_r^{\top} \mathbf{x}(t_k) - 
\mathcal{F}\Big(\mathcal{G}(\mathbf{s}_i(t_k), \dots, \mathbf{s}_i(t_{k-l}))\Big) \right\|_2^2.
\end{split}
\end{equation}
During the training phase, the LSTM ($\mathcal{G}$) and SDN ($\mathcal{F}$) models are trained using the reduced representation $\hat{\mathbf{v}}(t_k)$ in the latent space. Therefore, the full state vector $\hat{\mathbf{x}}(t_k)$ must be reconstructed from this low-dimensional representation. This is achieved through the truncated SVD basis, yielding
\begin{equation}
    \hat{\mathbf{x}}(t_k) = \mathbf{U}_r \, \hat{\mathbf{v}}(t_k),
\end{equation}
This compressed formulation provides a key advantage of SHRED: by training in a reduced latent space, the ML network can be efficiently trained in just a few minutes on a standard personal computer, as demostrated in \cite{williams2024sensing}.

\section{Numerical Results}\label{sec:results}
\begin{figure}
    \centering
    \includegraphics[width=.9\linewidth]{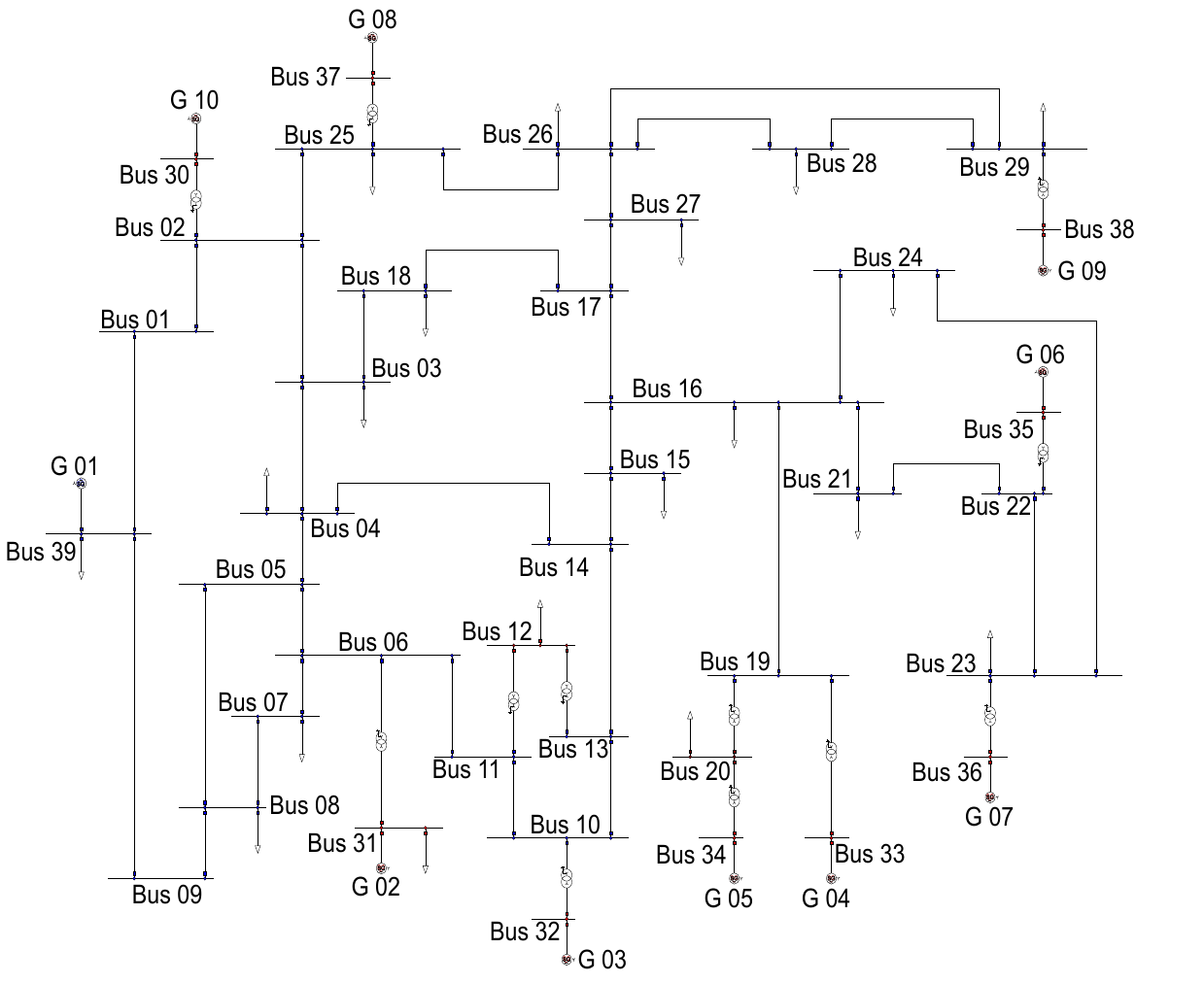}
    \caption{IEEE 39-Bus power system}
    \label{fig:IEEE39}
\end{figure}

This section presents a comprehensive analysis of SHRED applied to power system. All the analysis have been carried out on IEEE 39-Bus system, shown in Fig. \ref{fig:IEEE39}, which is widely used in the literature for DSE purpose. 

First, before presenting the case studies, a brief overview of the SHRED architecture is provided, as these remain consistent across all cases. SHRED is a supervised ML algorithm, and therefore requires training to learn the mapping from PMU measurements to the compressed representation of the full system state space. To generate the necessary dataset, dynamic simulations of the IEEE 39-bus power system were performed using DIgSILENT PowerFactory, a widely used commercial software for power system analysis \cite{gonzalez2014powerfactory}. The simulations span 15 different network configurations and consider line faults with clearing times ranging from 20~ms to 200~ms, thereby promoting a high degree of scenario diversity and generalization capability. The resulting dataset comprises 680 dynamic simulations, which are randomly divided into 80\% for training, 10\% for validation, and 10\% for testing. However, detailed information regarding the code used in this work are openly available through the GitHub repository referenced at the end of the paper. As discussed in Section~\ref{sec:SHRED}, the original architecture proposed in \cite{williams2024sensing} has been shown to perform effectively across a wide range of dynamical systems. Accordingly, the same architecture is adopted in this work. The LSTM component consists of two hidden layers with 64 neurons each, while the SDN component includes two hidden layers with 350 and 400 neurons, respectively \cite{ye2025pyshred}.\\
With respect to the problem formulation, Table~\ref{tab:inputoutputSHRED} summarizes the inputs and outputs associated with each snapshot used by the SHRED algorithm in the considered power system case study.
\begin{table}[htbp]
\caption{Inputs and Outputs of the proposed SHRED architecture.}
\label{tab:inputoutputSHRED}
\centering
\begin{tabular}{c||c|c|c}
\toprule
\textbf{Model} 
& \textbf{Inputs} 
& \multicolumn{2}{c}{\textbf{Outputs}} \\
\midrule
 & \textbf{PMUs} & \textbf{Buses} & \textbf{Generators} \\
\midrule
SHRED 
& $V,\, f,\, \theta$ 
& $V,\, f,\, \theta$ 
& $\omega,\, \delta,\, \psi_{1d},\, \psi_{1q},\, \psi_{2q},\, \psi_{\mathrm{exc}}$ \\
\bottomrule
\end{tabular}
\end{table}
As inputs, each PMU is assumed to provide voltage magnitude $V$ (p.u.), frequency $f$ (p.u.) and voltage angle $\theta$ (degrees) measurements at its corresponding bus, since not all PMUs are installed at line terminals and, therefore, active and reactive power measurements may not always be available.\\
Regarding the outputs, in the RMS simulations performed with DIgSILENT PowerFactory, synchronous generators are modeled using a sixth-order model\footnote{For further details on the dynamic model of synchronous generators, the reader is referred to the DIgSILENT PowerFactory manual \cite{gonzalez2014powerfactory}.}. The corresponding state variables include the rotor speed $\omega$ (p.u.), the rotor electrical angle $\delta$ (degrees), the flux in the $1d$ damper winding $\psi_{1d}$ (p.u.), the flux in the $1q$ damper winding $\psi_{1q}$ (p.u.), the flux in the $2q$ damper winding $\psi_{2q}$ (p.u.), and the excitation flux $\psi_{\mathrm{exc}}$ (p.u.). Moreover, SHRED is also trained to reconstruct $V$, $f$ and $\theta$ at all the buses, even though they are not generator states. \\
It is worth emphasizing that no direct comparison is provided against the KF, EKF, or UKF. These model-based approaches have demonstrated good performance for DSE in previous studies \cite{zhao2016robust,qi2016dynamic,zhao2017robust,zhao2017dynamic,hou2022dynamic}, provided that PMUs are optimally placed to ensure full system observability. However, if this condition is not satisfied, the performance of these estimators can significantly deteriorate and, in the absence of observability, state estimation may become infeasible \cite{qi2014optimal}.
The objective of this work is therefore not to outperform model-based estimators under ideal measurement configurations, but rather to address the gap between such assumptions and real-world operating conditions. In practice, PMUs are often sparsely deployed and rarely installed at generator terminals, making full observability difficult to achieve. In contrast, SHRED is specifically designed to perform DSE using a highly limited number of measurements and under non-optimal PMU placements, scenarios in which conventional model-based approaches may become infeasible. 

\subsection{Sensitivity analysis with respect to the number of PMUs}\label{sec:numberPMU}
In this section, a sensitivity analysis is carried out to evaluate how the number of PMUs affects SHRED ability to reconstruct the full system state space. To demonstrate the effectiveness of SHRED in power system applications, several case studies are analyzed in this work. These case studies, summarized in Table~\ref{tab:differentPMUs}, consider different numbers of PMUs deployed across the network.
 
\begin{figure}
    \centering
    \includegraphics[width=.9\linewidth]{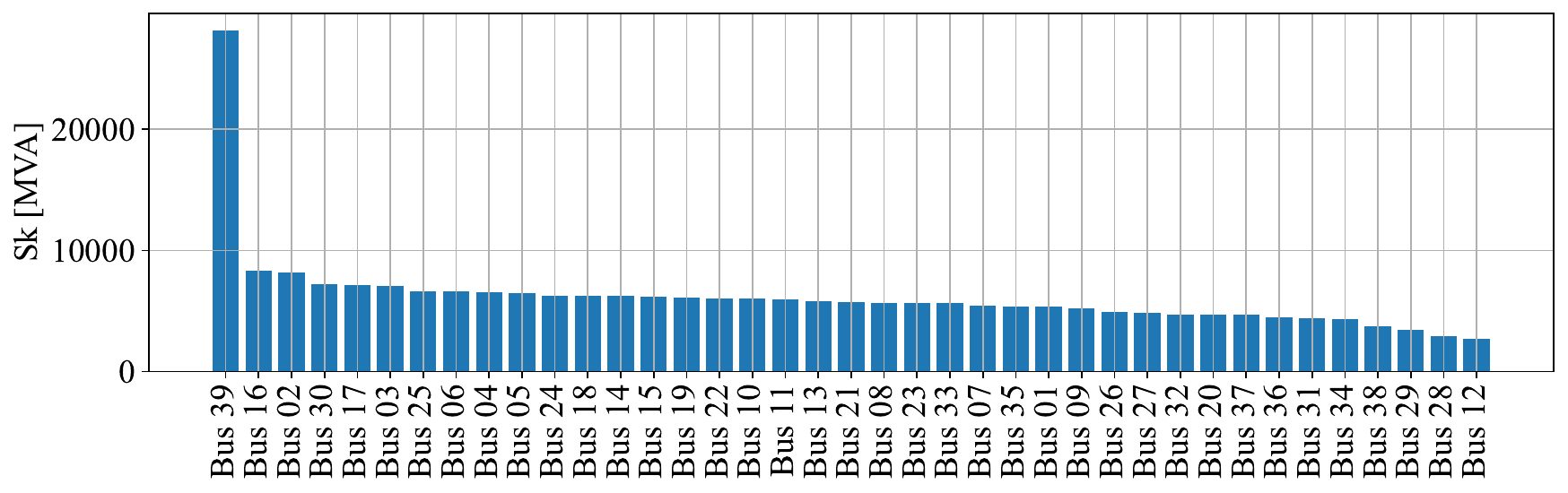}
    \caption{Short-circuit power at all the 39 buses}
    \label{fig:Scc}
\end{figure}

\begin{figure*}
    \centering
    \includegraphics[width=1\linewidth]{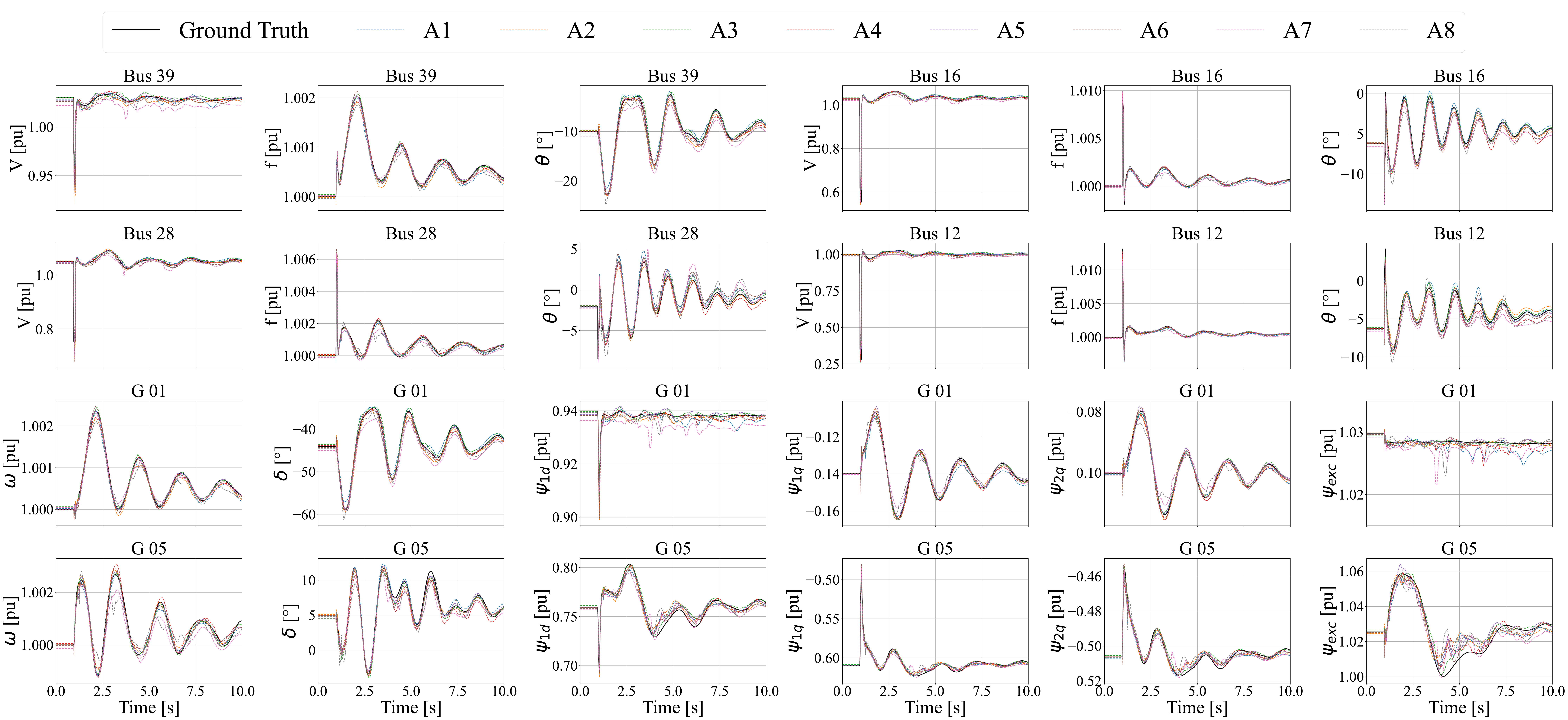}
    \caption{Reconstructed $V$, $f$ and $\theta$ at \textit{buses} $12$, $16$, $28$, and $39$, and generator state variables $\omega,\, \delta,\, \psi_{1d},\, \psi_{1q},\, \psi_{2q}$, and $\psi_{\mathrm{exc}}$ for $G01$ and $G05$ under a short-circuit test for all case studies with varying numbers of PMUs listed in Table~\ref{tab:differentPMUs}. SHRED demonstrates very high accuracy for case studies $A1$ through $A6$, with a slight decrease in reconstruction performance for case studies $A7$ and $A8$.}
    \label{fig:recon_1}
\end{figure*}

\begin{table}[htbp]
\caption{Case studies with different numbers of PMUs}
\label{tab:differentPMUs}
\centering
\begin{tabular}{c|c}
\toprule
\textbf{Case study} & \textbf{PMU deployment} \\
\midrule
A1 & PMUs installed at all 39 buses \\
A2 & PMUs installed at the 30 buses with the highest $S_{cc}$ \\
A3 & PMUs installed at the 20 buses with the highest $S_{cc}$ \\
A4 & PMUs installed at the 10 buses with the highest $S_{cc}$ \\
A5  & PMUs installed at the 5 buses with the highest $S_{cc}$ \\
A6  & PMUs installed at the 3 buses with the highest $S_{cc}$ \\
A7  & PMUs installed at the 2 buses with the highest $S_{cc}$ \\
A8  & PMU installed at the bus with the highest $S_{cc}$ \\
\bottomrule
\end{tabular}
\end{table}
Eight case studies are analyzed, starting from an ideal scenario in which all buses are equipped with PMUs and progressively reducing the number of PMUs until only a single bus is monitored. The PMU locations are selected based on the short-circuit power $S_{cc}$, with PMUs installed at the buses exhibiting the highest $S_{cc}$. This criterion is adopted because buses with high short-circuit power are typically considered the most influential nodes in a power system. Fig.~\ref{fig:Scc} illustrates the values $S_{cc}$ at all buses of the system.
To evaluate SHRED ability to reconstruct power system dynamics, all tests were carried out using fault scenarios that were not included in the training or validation datasets, thereby assessing the model capability to generalize to unseen dynamics. Figure~\ref{fig:recon_1} presents the SHRED reconstruction results for a short-circuit event. Due to space limitations, only the $V$, $f$, and $\theta$ at buses with the highest (\textit{Bus 39} and \textit{16}) and lowest (\textit{Bus 12} and \textit{28}) $S_{cc}$ are shown. For generator dynamics, the figure displays the reconstructed state variables of two representative machines: generator $G01$, which has the highest nominal power, and generator $G05$, which has the lowest nominal power.
Regarding the reconstruction of $V$, $f$, and $\theta$, SHRED demonstrates excellent performance across nearly all case studies. For the generator state variables, SHRED achieves very good accuracy for case studies $A1$ through $A6$. In contrast, in case studies $A7$ and $A8$, the reconstruction accuracy decreases slightly: specifically, $\psi_{1d}$ and $\psi_{\mathrm{exc}}$ for generator $G01$, and $\omega$ for generator $G05$ show larger deviations from the ground truth.

\begin{figure}
    \centering
    \includegraphics[width=1\linewidth]{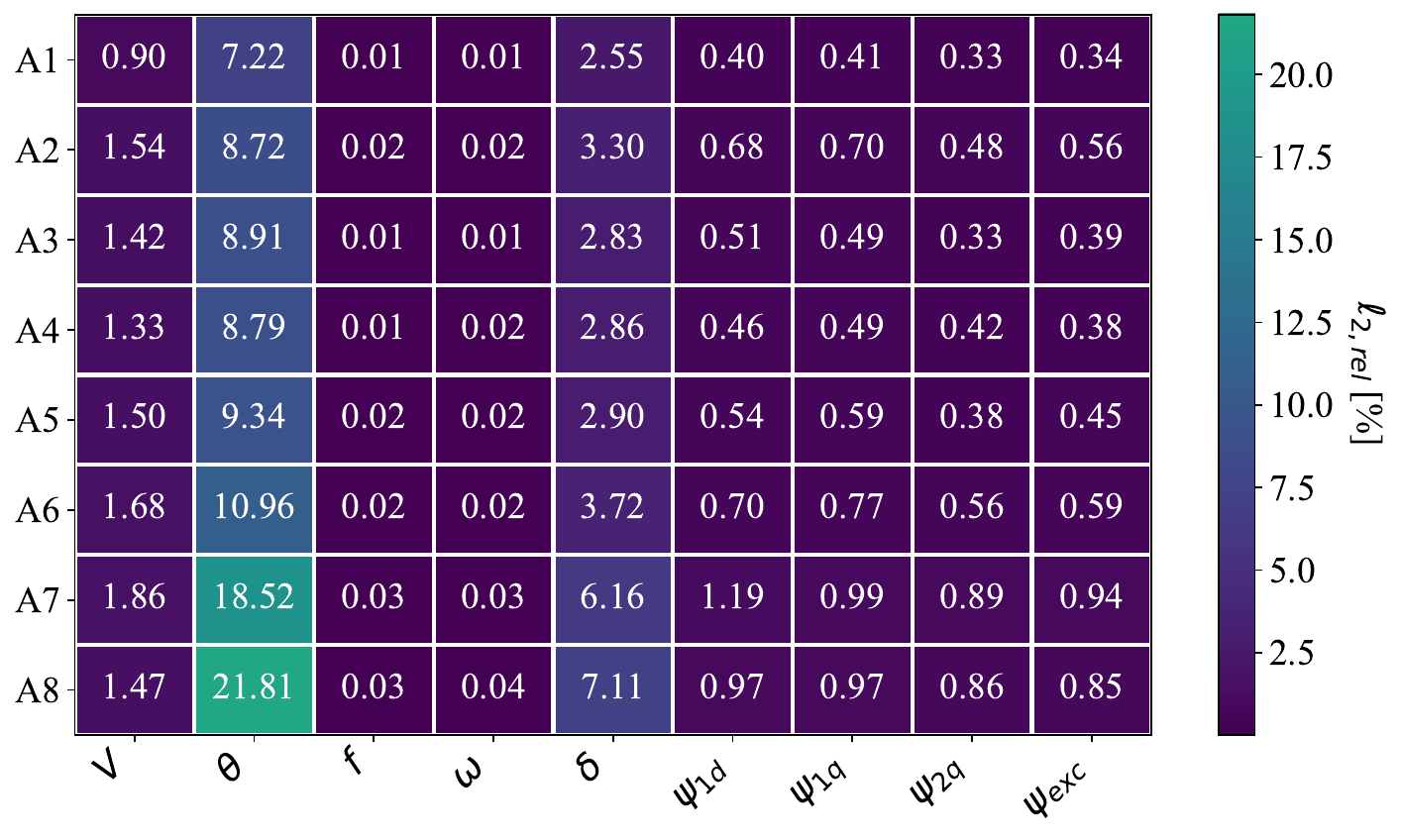}
    \caption{$\ell_{2,rel}[\%]$ values for each variable class across the case studies listed in Table~\ref{tab:differentPMUs}.}
    \label{fig:heatmap_A}
\end{figure}

To provide a comprehensive assessment of SHRED performance over the entire test dataset, the relative $\ell_2$ norm for each $i$-th output variable, denoted as $\ell_{2,rel}^{(i)}$, is defined as
\begin{equation}\label{eq:l2norm}
    \ell_{2,rel}^{(i)}[\%]= \frac{\left\| \mathbf{x}_{\text{true}}^{(i)} - \mathbf{\hat{x}}_{\text{SHRED}}^{(i)} \right\|_2}
    {\left\| \mathbf{x}_{\text{true}}^{(i)} \right\|_2}\cdot100,
\end{equation}
Here, $\mathbf{x}_{\text{true}}^{(i)}$ represents the ground-truth trajectories obtained from DIgSILENT PowerFactory simulations, while $\mathbf{\hat{x}}_{\text{SHRED}}^{(i)}$ denotes the corresponding SHRED reconstruction. The results are presented as a heatmap in Fig.~\ref{fig:heatmap_A}.
As observed, the largest reconstruction errors of SHRED occur for $\theta$, mainly because this variable strongly depends on the network configuration. However, across the entire test dataset, SHRED demonstrates excellent reconstruction performance, achieving a $\ell_{2,rel}$ error below 1\% for the variables $f$, $\omega$, $\psi_{1d}$, $\psi_{1q}$, $\psi_{2q}$, and $\psi_{\mathrm{exc}}$. For bus voltage magnitudes $V$, the relative $\ell_{2,rel}$ error remains below 2\% across all case studies. For $\delta$, the $\ell_{2,rel}$ error remains below 4\% as long as at least three PMUs are used as inputs. To provide further clarity, Table~\ref{tab:MeanErrorCases_A} reports the average $\ell_{2,rel}$ value for each case study.
\begin{table}[htbp]
\caption{Mean $\ell_{2,rel}[\%]$ for each case study listed in Tab. \ref{tab:differentPMUs}.}
\label{tab:MeanErrorCases_A}
\centering
\resizebox{\columnwidth}{!}{
\begin{tabular}{c|cccccccc}
\toprule
\toprule
& A1 & A2 & A3 & A4 & A5 & A6 & A7 & A8 \\
\midrule
Mean $\ell_{2,rel}[\%]$ 
& 1.352 
& 1.780 
& 1.656 
& 1.640 
& 1.749 
& 2.113 
& 3.401 
& 3.790 \\

\bottomrule
\bottomrule
\end{tabular}
}
\end{table}
As can be observed, the average $\ell_{2,rel}$ remains below 2\% for case studies $A1$ through $A5$, while it increases for the subsequent configurations. As the number of PMUs decreases, SHRED reconstruction accuracy gradually deteriorates. Nevertheless, the average error over the entire test dataset remains acceptable for all case studies. In particular, case studies $A7$ and $A8$ achieve a $\ell_{2,rel}$ below 4\% using only two and one PMU as input, respectively, representing an extremely challenging scenario for DSE.

\begin{figure}
    \centering
    \includegraphics[width=.8\linewidth]{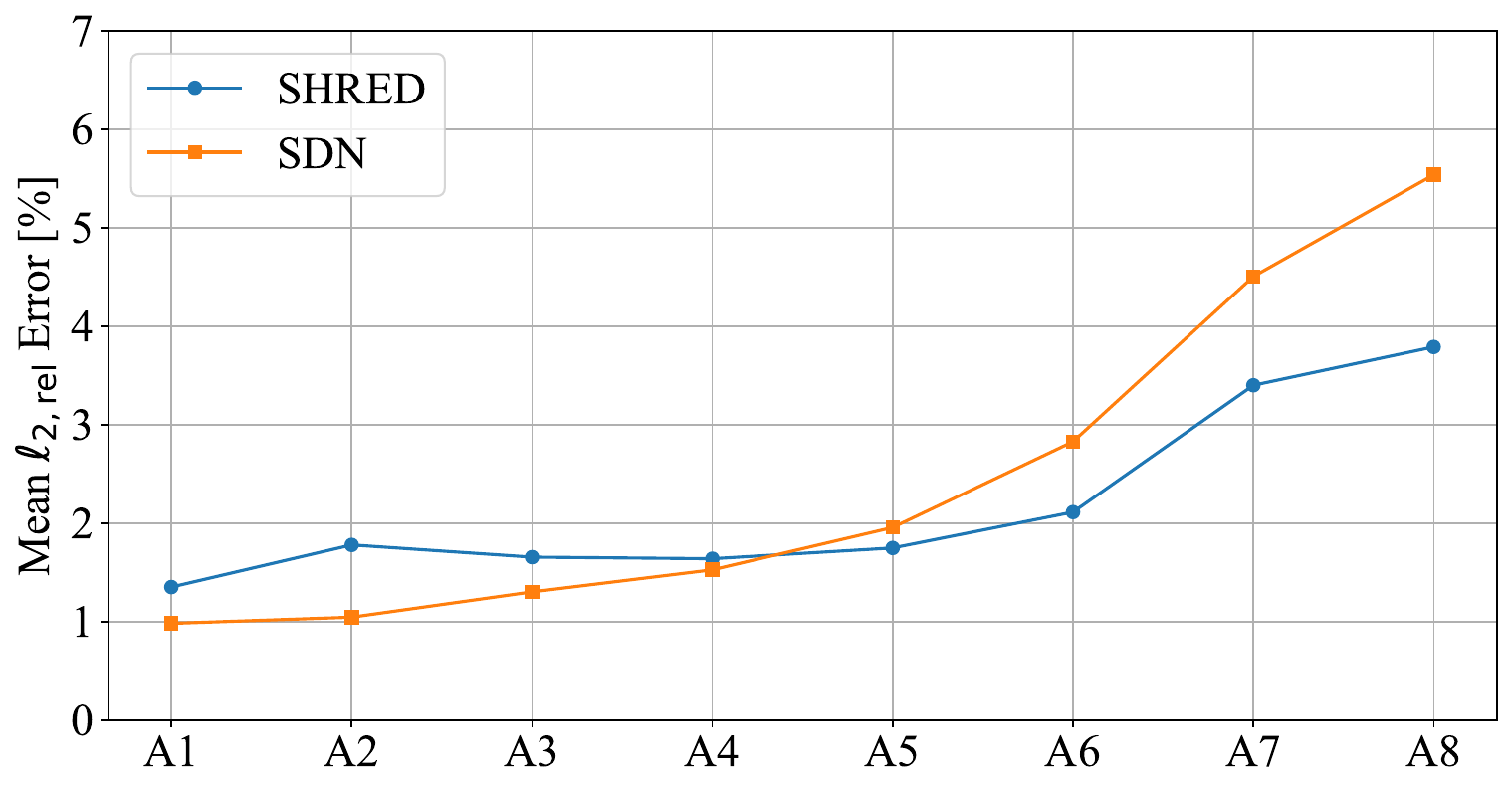}
    \caption{Comparison of the mean $\ell_{2,rel} [\%]$ values between SHRED and SDN for all the case studies in Tab. \ref{tab:differentPMUs}.}
    \label{fig:comparison_SDN}
\end{figure}

It is important to assess the performance of the proposed SHRED approach with respect to the state-of-the-art machine learning method. To assess the benefits of the proposed architecture, a comparison is performed against a baseline SDN. To assess this, Fig.~\ref{fig:comparison_SDN} reports the mean relative $\ell_2$ error of SHRED and SDN across all the case studies listed in Table~\ref{tab:differentPMUs}.

As can be observed, the advantage of SHRED over SDN becomes more pronounced when a limited number of measurements is available. When many measurements are accessible, both methods can effectively reconstruct the full state, as the mapping from measurements to the state space is less challenging. However, as the number of PMUs decreases, the performance of SDN degrades, whereas SHRED maintains higher accuracy. This behavior can be attributed to the LSTM-based temporal encoder in SHRED, which exploits the temporal history of the measurements to construct a more informative latent representation (Takens’ theorem \cite{takens2006detecting}).

\subsection{Sensitivity analysis with respect to the location of PMUs}\label{sec:locationPMU}
This section aims to demonstrate that SHRED is largely agnostic to PMU placement, which represents a significant advantage for power system applications. In practice, TSOs deploy PMUs according to different operational objectives, such as monitoring electromechanical oscillations, assessing voltage criticality, or enhancing situational awareness in specific areas of the network. As a result, PMU locations are often not optimized for model-based DSE approaches, for which suboptimal placement may lead to loss of observability and render DSE infeasible. To assess the robustness of SHRED with respect to PMU placement, nine different clusters of possible PMU locations are considered and evaluated. These configurations are summarized in Table~\ref{tab:locationPMUs}. 

\begin{table}[htbp]
\caption{Clusters with different PMU placement}
\label{tab:locationPMUs}
\centering
\begin{tabular}{c|c}
\toprule
\textbf{Cluster} & \textbf{Bus with installed a PMU} \\
\cmidrule(lr){1-1} \cmidrule(lr){2-2} 
B1 & 01, 04, 08, 11, 12, 16, 18, 19, 20, 25 \\
B2 & 02, 03, 05, 09, 13, 14, 22, 23, 27, 28\\
B3 & 06, 07, 10, 15, 17, 21, 24, 26, 29, 39\\
\cmidrule(lr){1-1} \cmidrule(lr){2-2}
B4 & 01, 12, 13, 14, 16\\
B5 & 02, 09, 11, 18, 28\\
B6 & 08, 19, 20, 24, 39\\
B7 & 05, 07, 10, 17, 27\\
B8 & 03, 04, 22, 23, 29\\
B9 & 06, 15, 21, 25, 26\\
\bottomrule
\end{tabular}
\end{table}
All clusters consider PMU installations exclusively at High-Voltage buses. Medium-Voltage generator buses are not included, as these buses typically belong to power plant owners and are therefore not directly accessible for PMU installation by the TSO.  
Clusters $B1$ to $B3$ are constructed by randomly placing 10 PMUs across the HV network, whereas clusters $B4$ to $B9$ consider random placements of 5 PMUs. The choice of configurations with 10 and 5 PMUs, corresponding to approximately 33\% and 17\% of the HV buses, respectively, is motivated by realism: in current WAMS, less than 50\% of transmission-level buses are typically equipped with PMUs.

\begin{figure*}
    \centering
    \includegraphics[width=1\linewidth]{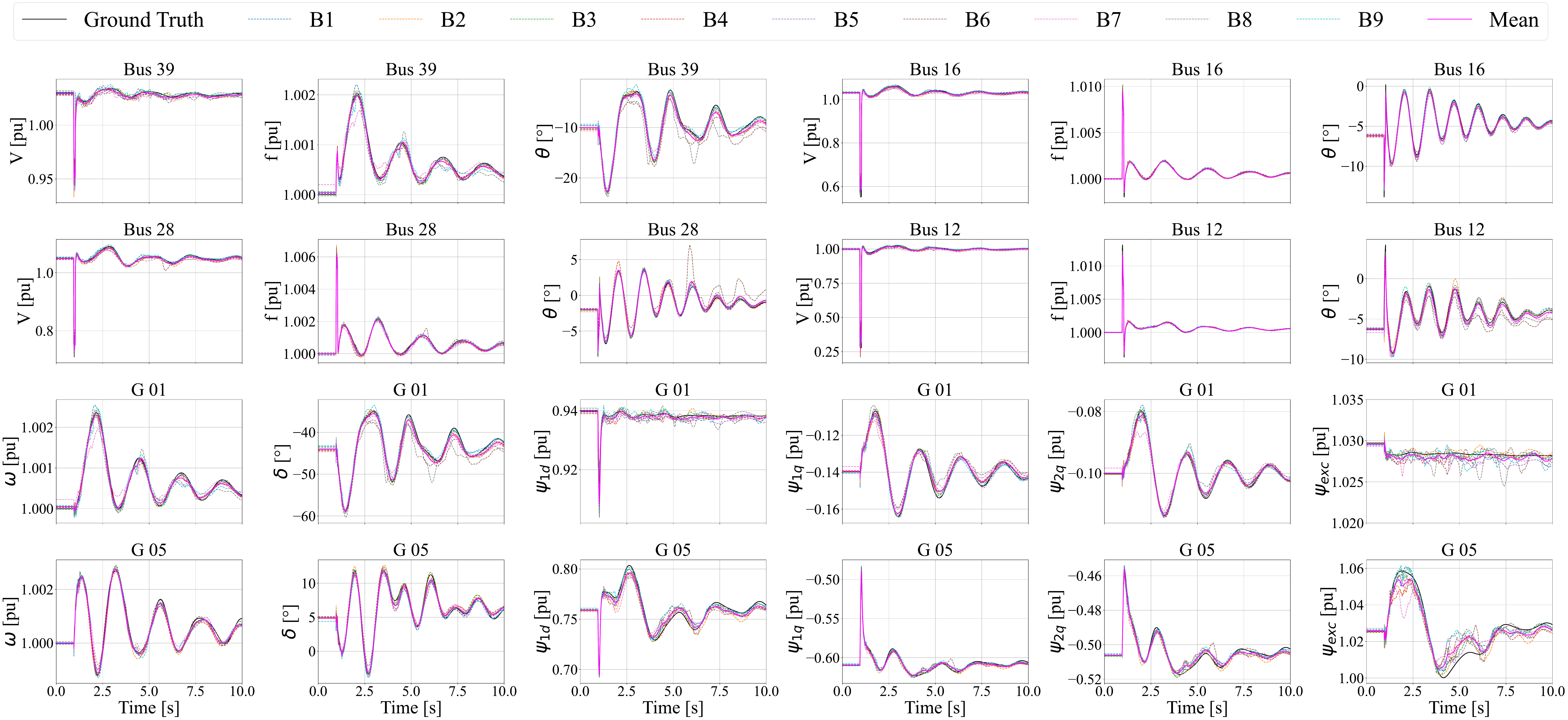}
    \caption{Reconstructed $V$, $f$, and $\theta$ at \textit{buses} $12$, $16$, $28$, and $39$, along with the generator state variables $\omega$, $\delta$, $\psi_{1d}$, $\psi_{1q}$, $\psi_{2q}$, and $\psi_{\mathrm{exc}}$ for generators $G01$ and $G05$ under a short-circuit test for all case studies with varying PMU placements, as listed in Table~\ref{tab:locationPMUs}. SHRED achieves consistent reconstruction performance across all clusters, with the mean reconstruction over all clusters exhibiting higher robustness and accuracy than individual cluster results.}
    \label{fig:recon_2}
\end{figure*}

To visualize SHRED performance under different PMU placement configurations, the same short-circuit event considered in the previous section is simulated, and the corresponding results are presented in Fig.~\ref{fig:recon_2}. In addition, in this figure, the mean reconstruction across all clusters is reported, as it provides a more robust estimate with respect to variations in PMU placement.

\begin{figure}
    \centering
    \includegraphics[width=1\linewidth]{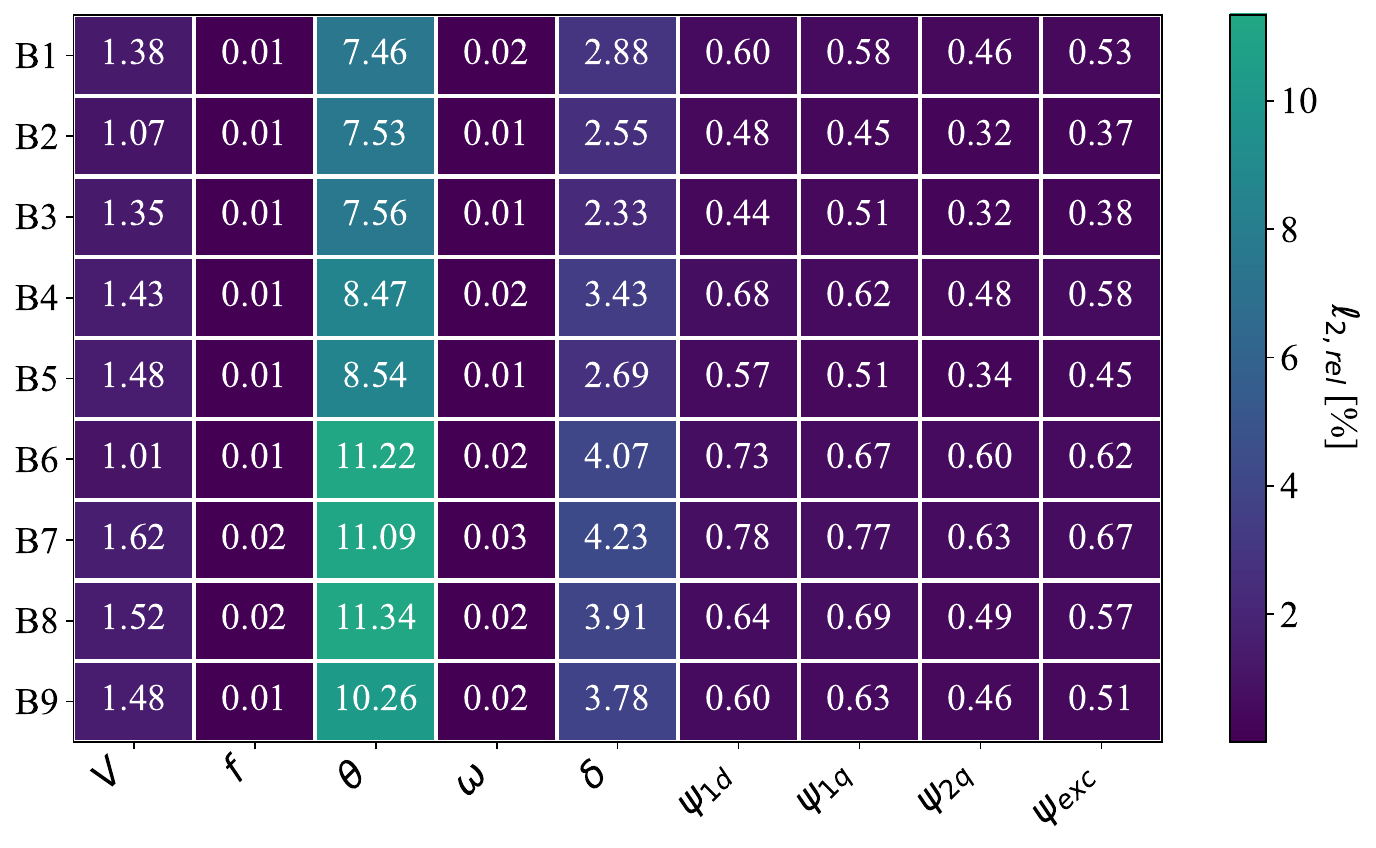}
    \caption{$\ell_{2,rel} [\%]$ values for each variable class across the case studies listed in Table~\ref{tab:locationPMUs}.}
    \label{fig:heatmap_B}
\end{figure}

As can be observed, all the considered clusters are able to accurately reconstruct the power system dynamics, including both $V$, $f$, and $\theta$ at the buses and the dynamic state variables $\omega$, $\delta$, $\psi_{1d}$, $\psi_{1q}$, $\psi_{2q}$, and $\psi_{\mathrm{exc}}$ of the generators. Interestingly, the mean reconstruction obtained across all clusters exhibits better performance than each individual cluster. This behavior can be explained by the fact that individual clusters may present localized reconstruction errors at specific time instants, while averaging the results across clusters mitigates such errors, leading to a more robust overall reconstruction.\\
Furthermore, the $\ell_{2,rel}$, introduced in \eqref{eq:l2norm}, is adopted to provide a comprehensive overview and a clear quantitative indicator of SHRED performance across all the clusters on all the test dataset. Figure~\ref{fig:heatmap_B} presents the corresponding heatmap, enabling identification of the variable classes associated with the largest reconstruction errors.
As can be inferred, SHRED worst performance remains on the $\theta$, consistent with the observations in the previous analysis. It is worth noting that across the different PMU clusters, the reconstruction error for each variable is remarkably similar, highlighting SHRED robustness to variations in PMU placement. This is further confirmed in Tab.~\ref{tab:MeanErrorCases_B}, where the average $\ell_{2,rel}$ is reported for each cluster, showing that SHRED maintains similar performance across different PMU placements.

\begin{table}[htbp]
\caption{Mean $\ell_{2,rel}[\%]$ for each cluster listed in Tab. \ref{tab:locationPMUs}.}
\label{tab:MeanErrorCases_B}
\centering
\resizebox{\columnwidth}{!}{
\begin{tabular}{c|ccccccccc}
\toprule
\toprule
& B1 & B2 & B3 & B4 & B5 & B6 & B7 & B8 & B9\\
\midrule
Mean $\ell_{2,rel}[\%]$ 
& 1.546 
& 1.421
& 1.434 
& 1.747 
& 1.622 
& 2.106 
& 2.204
& 2.133 
& 1.972\\

\bottomrule
\bottomrule
\end{tabular}
}
\end{table}

\subsection{Numerical analysis with measurement noise}\label{sec:noise}
\begin{figure*}
    \centering
    \includegraphics[width=1\linewidth]{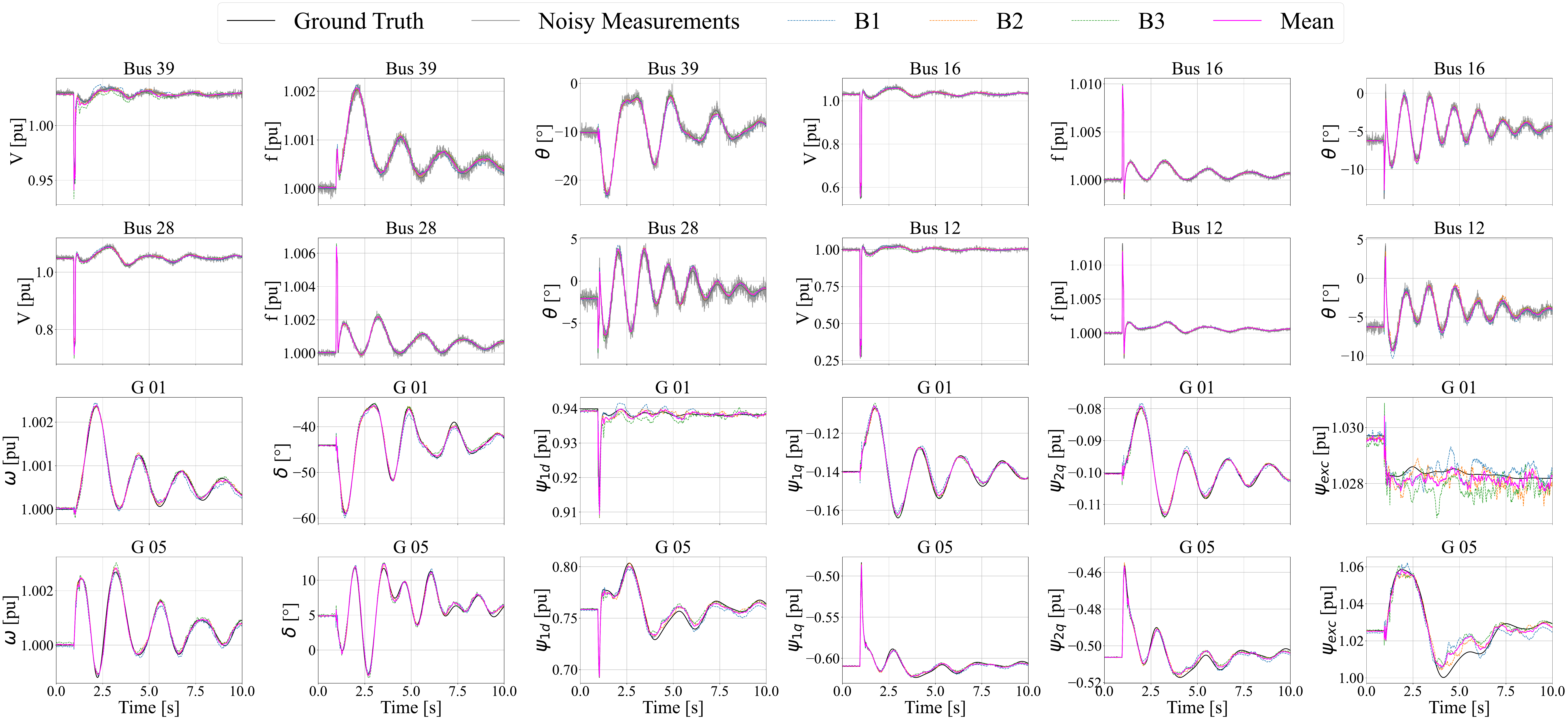}
    \caption{Reconstructed $V$, $f$, and $\theta$ at \textit{buses} $12$, $16$, $28$, and $39$, along with the generator state variables $\omega$, $\delta$, $\psi_{1d}$, $\psi_{1q}$, $\psi_{2q}$, and $\psi_{\mathrm{exc}}$ for generators $G01$ and $G05$ under a short-circuit test for case studies $B1$, $B2$ and $B3$, listed in Table~\ref{tab:locationPMUs}, adding measurement noise to the PMUs. SHRED achieves consistent reconstruction performance across all clusters even in presence of measurement noise.}
    \label{fig:recon3_noisy}
\end{figure*}

This section investigates the performance of SHRED under measurement noise, which is a common characteristic of real-world power system data. To this end, we consider the first three PMU placement clusters listed in Tab.~\ref{tab:locationPMUs} ($B1$, $B2$, and $B3$) and corrupt the PMU measurements with Gaussian noise as follows:
\begin{equation}
    s_{i,\text{noisy}} = s_i + \mathcal{N}(0, \sigma_{s_i}^2)
\end{equation}
where $s_i$ represents the ground truth, and $\mathcal{N}$ denotes a Gaussian distribution with zero mean and variance $\sigma_{s_i}^2$. In this work, the noise variance is defined as a function of the standard deviation of the original measurements, according to
\begin{equation}
    \sigma_{s_i} = \alpha \cdot \text{std}\{s_i\}
\end{equation}
where $\alpha$ denotes the noise level and $\text{std}\{\cdot\}$ represents the standard deviation of the corresponding measurement signal. Different values of $\alpha$ were investigated; however, for sake of compactness, only the case with $\alpha = 0.15$ is reported here, which corresponds to a relatively high level of measurement noise contamination. The same short-circuit event considered in the previous sections is used and Fig.~\ref{fig:recon3_noisy} shows the reconstruction for $V$, $f$, and $\theta$ at \textit{buses} $12$, $16$, $28$, and $39$ compared with the ground truth and the noisy PMU measurements provided as input. Furthermore, it presents the reconstruction of the generator state variables $\omega$, $\delta$, $\psi_{1d}$, $\psi_{1q}$, $\psi_{2q}$, and $\psi_{\mathrm{exc}}$ for generators $G01$ and $G05$. It is worth noting that no noisy measurements are shown for the generator states, since these variables cannot be directly measured in real-world.

\begin{figure}
    \centering
    \includegraphics[width=1\linewidth]{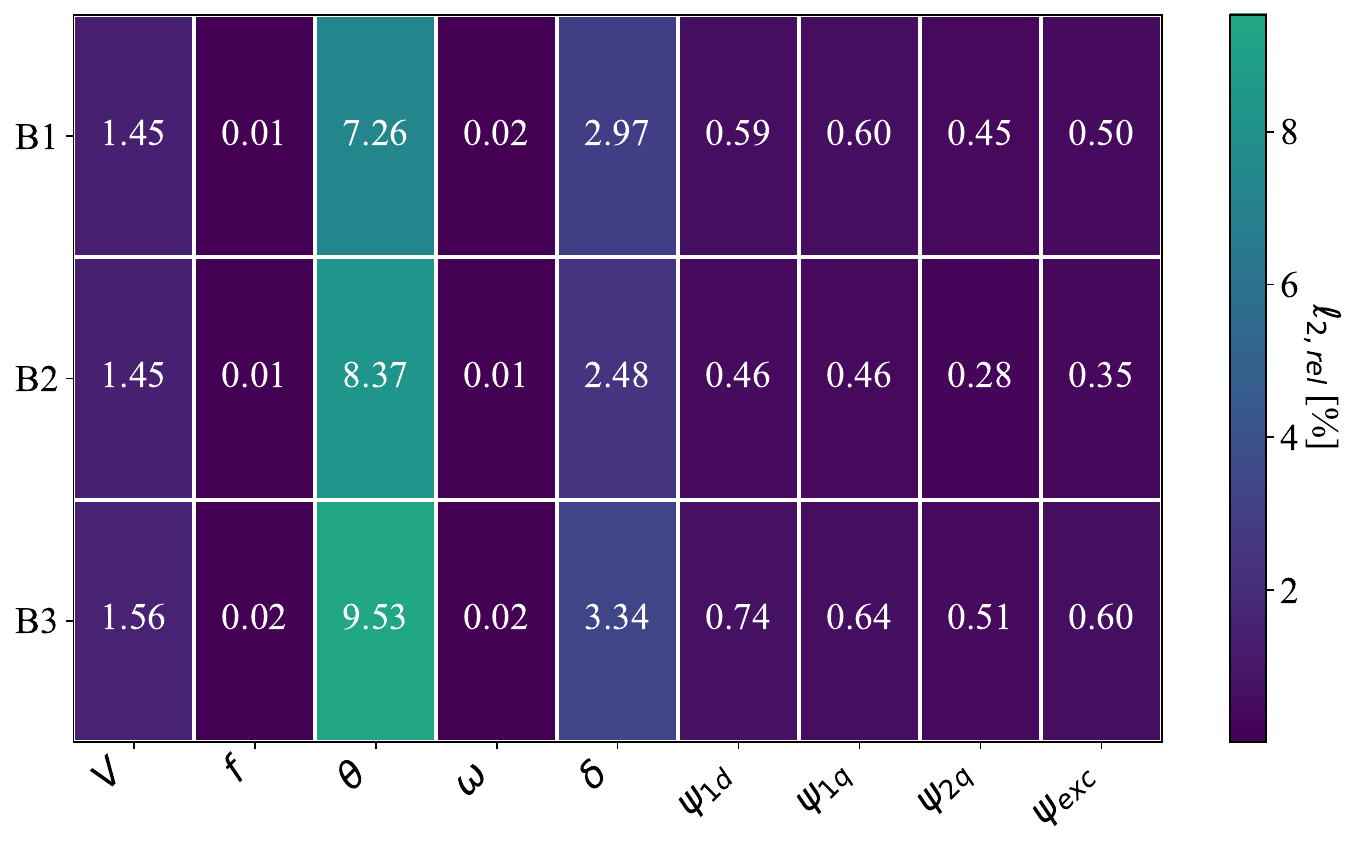}
    \caption{$\ell_{2,rel} [\%]$ values for each variable class across the case studies B1, B2 and B3, listed in Table~\ref{tab:locationPMUs}, with measurements noise.}
    \label{fig:heatmap_L2rel_noisy}
\end{figure}

As can be observed, the overall dynamic trend remains consistent with the results obtained in the noise-free case. In particular, the mean reconstruction, computed as the ensemble average of all SHRED models, demonstrates improved smoothness and robustness, maintaining accurate performance even under significant measurement noise. To further quantify this behavior, Fig.~\ref{fig:heatmap_L2rel_noisy} reports the corresponding $\ell_{2,rel}$ values for the three considered clusters.

As can be observed by comparing these results with the noise-free case in Fig. \ref{fig:heatmap_B}, SHRED reconstructing performance of $V$, $f$ and $\theta$ slightly deteriorates in the presence of measurement noise. Lastly, Table \ref{tab:MeanErrorCases_C} reports the average $\ell_{2,rel}[\%]$ of three case studies. 

\begin{table}[htbp]
\caption{Mean $\ell_{2,rel}[\%]$ for cluster $B1$, $B2$ and $B3$ adding measurement noise.}
\label{tab:MeanErrorCases_C}
\centering

\begin{tabular}{c|ccc}
\toprule
\toprule
& B1 & B2 & B3 \\
\midrule
Mean $\ell_{2,rel}[\%]$ 
& 1.539
& 1.541
& 1.844 
\\

\bottomrule
\bottomrule
\end{tabular}

\end{table}

As can be observed, the obtained values are comparable to those reported for the noise-free case in Table~\ref{tab:MeanErrorCases_B}. Therefore, the $\ell_{2,rel}$ values remain low, confirming that SHRED is capable of accurately reconstructing system dynamics even in the presence of measurement noise and strongly nonlinear disturbances, such as short-circuit events.

Lastly, it is worth discussing the computational requirements of SHRED for both training and inference. All experiments were performed on an NVIDIA T600 GPU. Across all case studies reported in Tables~\ref{tab:differentPMUs} and \ref{tab:locationPMUs}, the average training time was 13:27~min, ranging from a minimum of 11:57~min to a maximum of 17:30~min. However, the training stage is performed offline and therefore does not represent a major constraint for real-time DSE applications. More relevant is the inference time, which averaged 1.3~ms, with values ranging between 0.6~ms and 3.1~ms on the same GPU. Considering that WAMS measurements are typically acquired at a sampling period of 20~ms in Europe, the proposed SHRED framework is sufficiently fast to perform DSE in real time.

\section{Limitations and Future Work}\label{sec:limitations}
The main drawback of SHRED, being a supervised machine learning algorithm, is the requirement of full state trajectories during the training phase, which are not measurable in real-world power systems. However, as demonstrated in \cite{bao2025data}, a Data Assimilation version of SHRED (DA-SHRED), trained on a high-fidelity model and subsequently deployed in a real dynamic system, is capable of accurate state reconstruction. This approach is plausible in the power system context, since TSOs already possess detailed dynamic models used for accurate DSA. Therefore, SHRED could be trained offline on validated system models and then deployed online using real PMU measurements, bridging the gap between simulation-based training and real-world operation.

Future work will focus on validating DA-SHRED on real-world power systems by training the framework on high-fidelity simulations and evaluating its ability to reconstruct the dynamic state during strongly nonlinear disturbances, such as short-circuit events. In addition, future research will investigate the integration of SHRED with SINDy to discover interpretable latent-space governing equations, following the framework proposed in \cite{gao2025sparse}. This direction may provide new insights into power system dynamics while combining accurate state reconstruction with physics-informed model discovery.

\section{Conclusions}\label{sec:conclusion}
This paper introduces SHRED, a novel machine learning framework for Dynamic State Estimation. Existing DSE approaches are predominantly based on Kalman filter variants or machine learning techniques. However, Kalman-based methods often suffer from high computational complexity, sensitivity to model inaccuracies, and performance degradation under strongly nonlinear operating conditions. Furthermore, their effectiveness critically depends on measurement availability and placement, as suboptimal PMU configurations can reduce observability and even render state estimation infeasible. On the other hand, existing machine learning approaches can alleviate some of these limitations, but they often require large training datasets and substantial computational resources. This can result in significant training costs and may hinder their scalability and practical deployment for large-scale power system applications.\\
To address these limitations, SHRED is proposed as a data-driven framework capable of reconstructing the full system state from a limited number of PMU measurements. The results demonstrate that SHRED maintains high estimation accuracy even when only sparse measurements are available and is largely insensitive to both the number and placement of PMUs. This represents a significant practical advantage for TSOs, since PMUs are typically installed to satisfy specific monitoring objectives, such as oscillation monitoring or voltage stability assessment, rather than to guarantee observability for model-based DSE algorithms. Consequently, measurement configurations that are inadequate for conventional Kalman-filter-based estimators can still provide accurate state reconstruction when using SHRED. In addition, the proposed framework has been shown to be robust to measurement noise and severe nonlinear disturbances, further supporting its applicability to real-world power system operation.

\section*{Code}

The code is available at: \href{https://github.com/Andrea-Pomarico/SHRED_PowerGridsDSE/tree/main}{\textcolor{blue}{SHRED-PowerGrids}}.

\bibliography{biblio.bib}
\bibliographystyle{IEEEtran}

\newpage

\vfill

\end{document}